\documentclass[]{spie}  
\usepackage[dvips]{graphicx}

\title{SkyDOT (Sky Database for Objects in the Time Domain):\\
A Virtual Observatory for Variability Studies at LANL}

\author{P.~Wozniak, K.~Borozdin, M.~Galassi, W.~Priedhorsky,
D.~Starr, W.~T. Vestrand, R.~White\\and\\ J.~Wren
\skiplinehalf
Los Alamos National Laboratory, Los Alamos, NM, USA
}

\authorinfo{Send correspondence to P.~Wozniak:\\
E-mail: wozniak@lanl.gov, Telephone: 1 505 667 1381,\\
Address: Los Alamos National Laboratory, MS-D436, Los Alamos, NM 87545, USA
}

\begin{document} 
  \maketitle 

\begin{abstract}

The mining of Virtual Observatories (VOs) is becoming a powerful new method for
discovery in astronomy. Here we report on the development of SkyDOT (Sky
Database for Objects in the Time domain), a new Virtual Observatory, which is
dedicated to the study of sky variability. The site will confederate a number
of massive variability surveys and enable exploration of the time domain in
astronomy. We discuss the architecture of the database and the functionality of
the user interface. An important aspect of SkyDOT is that it is continuously
updated in near real time so that users can access new observations in a timely manner.
The site will also utilize high level machine learning tools that will allow
sophisticated  mining of the archive. Another key feature is the real time
data stream provided by RAPTOR (RAPid Telescopes for Optical Response), a
new sky monitoring experiment under construction at Los Alamos National Laboratory (LANL).

\end{abstract}



\section{INTRODUCTION}
\label{sec:intro}  

The past decade astronomy has seen the advent of numerous
data intensive projects. In the early nineties, microlensing
experiments, which daily monitored tens of millions of objects over long
periods of time, were pushing the limits of commonly available computer storage
and processing power (e.g. Ref.~\citenum{Paczynski96}).
Typically microlensing teams implemented their own specialized
database systems with very limited portability. As a result, development effort was often
duplicated. Nevertheless, the scientific payoff of those projects went far
beyond the primary goal, that is the discovery and study of microlensing
events. The wealth of data created an information rich environment, where
serendipitous science happens continuously in studies of stellar populations
in the dense fields toward the Galactic Bulge and Magellanic Clouds\cite{Paczynski00}.

Microlensing searches provided an unprecedented record of variability
in those selected fields, with hundreds of epochs for each object gathered
over the several year baseline. However, no such record has been published
for most of the sky. In fact, the bright sky between 6 and 15 mag
is largely unexplored in terms of variability\cite{Paczynski97}.
There is even less data for astronomical events at short time scales,
that have to be identified in real time in order to be studied.
Automated online data analysis with alert capability is required
for success in this domain. The list of active projects measuring positions and brightness
over significant parts of the sky includes more than 30 names
({\tt http://www.astro.princeton.edu/faculty/bp.html}). Only a few of these
manage to process the data timely and make it available to all astronomers
in a useful form. Data overload seems to be a frequent occurrence
in astronomy today.

The concept of the Virtual Observatory\cite{NVO01} promises to solve many of the
problems with very large data sets. It is recognized that increasing
the amount of available data by an order of magnitude, and considering
joint multi-wavelength, spatial and temporal information from multiple
surveys simultaneously, opens up a new discovery space. Variability studies are
essential to numerous astronomical questions, however the time dimension
adds even more information to be processed. 
With the use of modern database systems and emerging standards for
web interfaces, numerous data sets can be federated, despite the fact that differing
technologies may be used at different nodes\cite{Williams02,Ochsenbein02}.
Data Mining and Machine Learning are becoming very attractive tools for
extracting knowledge from vast quantities of data.

The Sloan Digital Sky Survey (SDSS) is an example of a project where
technical issues of the efficient data distribution were given serious
consideration\cite{Szalay02}, although the SDSS SkyServer Database
does not provide the temporal data. There are several project that do have
time domain data and are working on making those available to the astronomical
community\cite{Brunner02}. To the best of our knowledge, none of
the teams (including the authors of this contribution) can provide the
full sky coverage with prompt online data access. This paper describes the work
in progress at the Los Alamos National Laboratory to build a Virtual
Observatory for studies of variable objects across the sky.

\section{Interactive Astronomical Variability Database at LANL}

   \begin{figure}[t]
   \begin{center}
   \begin{tabular}{c}
   \includegraphics[height=8cm, angle=0]{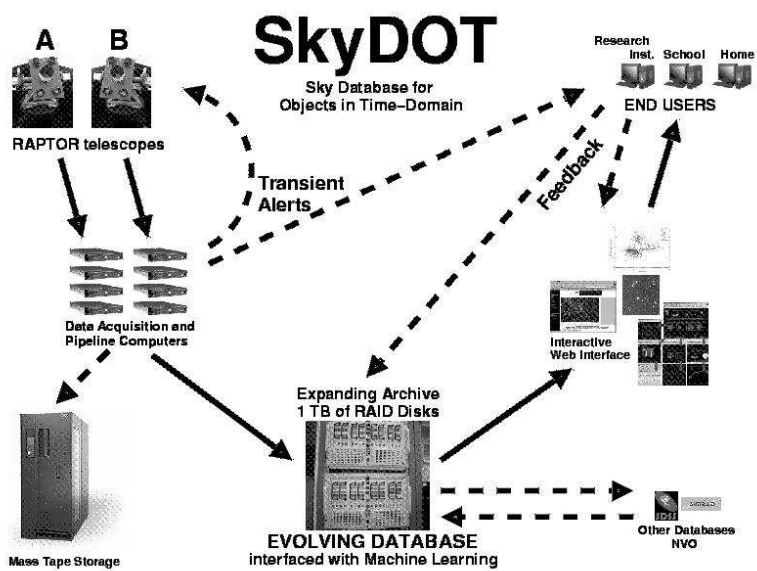}
   \end{tabular}
   \end{center}
   \caption[example] 
   { \label{fig:concept} The concept of SkyDOT, A Virtual Observatory for
Variability Studies at LANL. A real time data stream from RAPTOR will be
integrated with the database. The main data flow is clockwise from the upper
left to the upper right corner, that is from data acquisition hardware all the
way to the broad user. Tape backup and interfaces with other databases are also
shown. Important parts of the design are closed loop operation of RAPTOR
telescopes, real time transient alerts and feedback collecting.
}
   \end{figure} 


\subsection{Data Sources} 
\label{sec:sources}

The concept of the Virtual Observatory for studies of variable
objects is illustrated in Fig.~\ref{fig:concept}.
Los Alamos National Laboratory (LANL) is involved in several
sky monitoring projects. It is planned that the sky variability database we are
constructing will consist of multiple data sets, all converted to similar
formats and available through a common user interface with easy to apply
graphical tools. Some of the current data sets that will be included are
public OGLE-II data, RAPTOR and ROTSE.
One of the authors developed a Difference Image Analysis
pipeline for the Optical Gravitational Lensing Experiment and processed OGLE-II
Galactic Bulge data (Ref.~\citenum{Wozniak02a} and references therein).
RAPid Telescopes for Optical Response (RAPTOR) is a new, stereoscopic search
for optical transients in real time\cite{Vestrand01}. The Robotic
Optical Transient Search Experiment (ROTSE) is a GRB followup project
responding to satellite triggers\cite{Akerlof00a}.
Mirrors of other publicly available synoptic data
sets will also be incorporated over time. Below we summarize briefly what to
expect from each of the data sets.

\subsubsection{OGLE}
\label{sec:rotse}

OGLE-II survey\cite{Udalski97} was conducted with the 1.3 meter
Warsaw telescope at the
Las Campanas Observatory, Chile. The portion of the data collected during
observing seasons 1997--1999 has been analyzed using the Difference Image
Photometry, approximately calibrated to a standard system, and is available
in public domain\cite{Wozniak02a}. Only variable objects have been measured with
this technique. Between 200 and 300 $I$-band frames are available for each of
the 49 OGLE-II Galactic Bulge fields. The number of detected variable objects
per field varies between 800 and 9000 due to variations of the stellar density
and uneven frequency of observations. Each field covers $14\times57$ arcmin,
for a total of $\sim 11$ square deg. The range of covered galactic longitudes
is roughly $\pm11$ deg. The database comprises a total of over $51\times10^6$
individual photometric records for 221,801 objects with $I$-band magnitudes
between 10.5 and 20.0. The rate of spurious objects is still about 10\%.

\subsubsection{RAPTOR}
\label{sec:rotse}

RAPTOR\cite{Vestrand01,Vestrand02} is a new generation optical transient search
at LANL, NM. Its key features are real time data analysis, closed loop operation
with rapid slewing and response to interesting events, and stereo vision for
high confidence rejection of artifacts. Each of the two identical RAPTOR
telescopes, separated by 38 km, will have four 85 mm cameras with the 1500
square deg. total field of view, and a central, more sensitive 400 mm camera
with much narrower, $2\times2$ deg. field of view. An 0.3 m Ritchey-Chretien
telescope with a transmission grating will provide low resolution spectroscopy
for selected objects. The data rate from all imaging instruments will reach
4 TB/year. Fast cadence time histories will be constructed for roughly 300,000
objects across all locally visible sky up to the limiting magnitude 12.5.
With the additional sky patrol instrument, RAPTOR experiment can also cover all
locally visible sky to about 16 mag (about 30 million objects) in about 2--3
nights. The main challenge for the RAPTOR database will be real time operations.
All data will be available in public domain as soon as technically possible,
preferably as a real time update to the online database.

\subsubsection{ROTSE}
\label{sec:rotse}

For almost 4 years the ROTSE-I telescope nightly patrolled all the sky visible from
Los Alamos, NM. Those observations constitute a valuable database
for studying the variability of the sky in the 8--15.5 magnitude range.
For the purpose of an all-sky variability census\cite{Wozniak02b}
we are using the data taken between April 1999 and March 2000. Observations were
performed in 640 fields, each covering $8\times8$ deg. The most difficult part
of data processing, reducing images to object lists, is complete. This data set
amounts to more than 225,000 wide field images totaling approximately 2.5 TB
of data. Photometry alone takes 250 GB of binary storage. The number of available
observations varies between about 300 and 40
near declinations $+90$ deg and $-30$ deg respectively. Roughly 32,000 periodic
variables are expected to be found based on the scaling of a pilot study.
The total number of objects with time histories is over $2\times 10^7$,
and the total number of individual photometric measurements reaches
$3.5\times10^9$.

   \begin{figure}[h]
   \begin{center}
   \begin{tabular}{c}
   \includegraphics[height=11cm, angle=0]{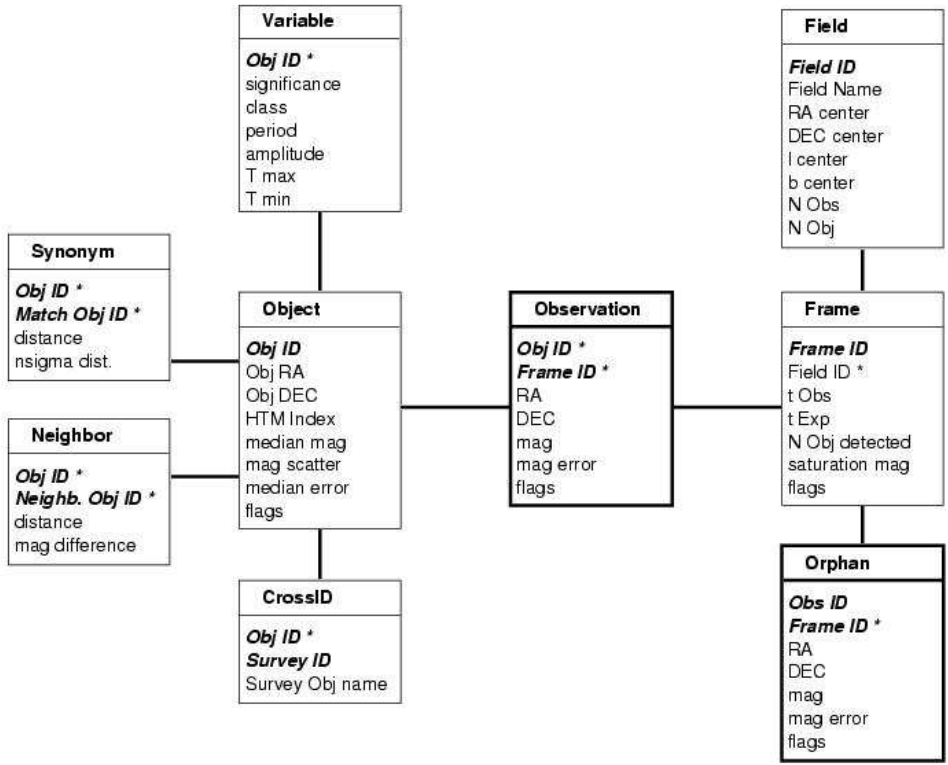}
   \end{tabular}
   \end{center}
   \caption[example] 
   { \label{fig:schema} The relational schema of the variability database
under construction. Event tables are shown with thick lines.
Bold slanted font indicates primary key attributes and asterisk indicates
a foreign key. Various parts of the snowflake surrounding the main fact
table serve as entry points for most queries.
}
   \end{figure} 

   \begin{figure}[h]
   \begin{center}
   \begin{tabular}{c}
   \includegraphics[height=7cm, angle=0]{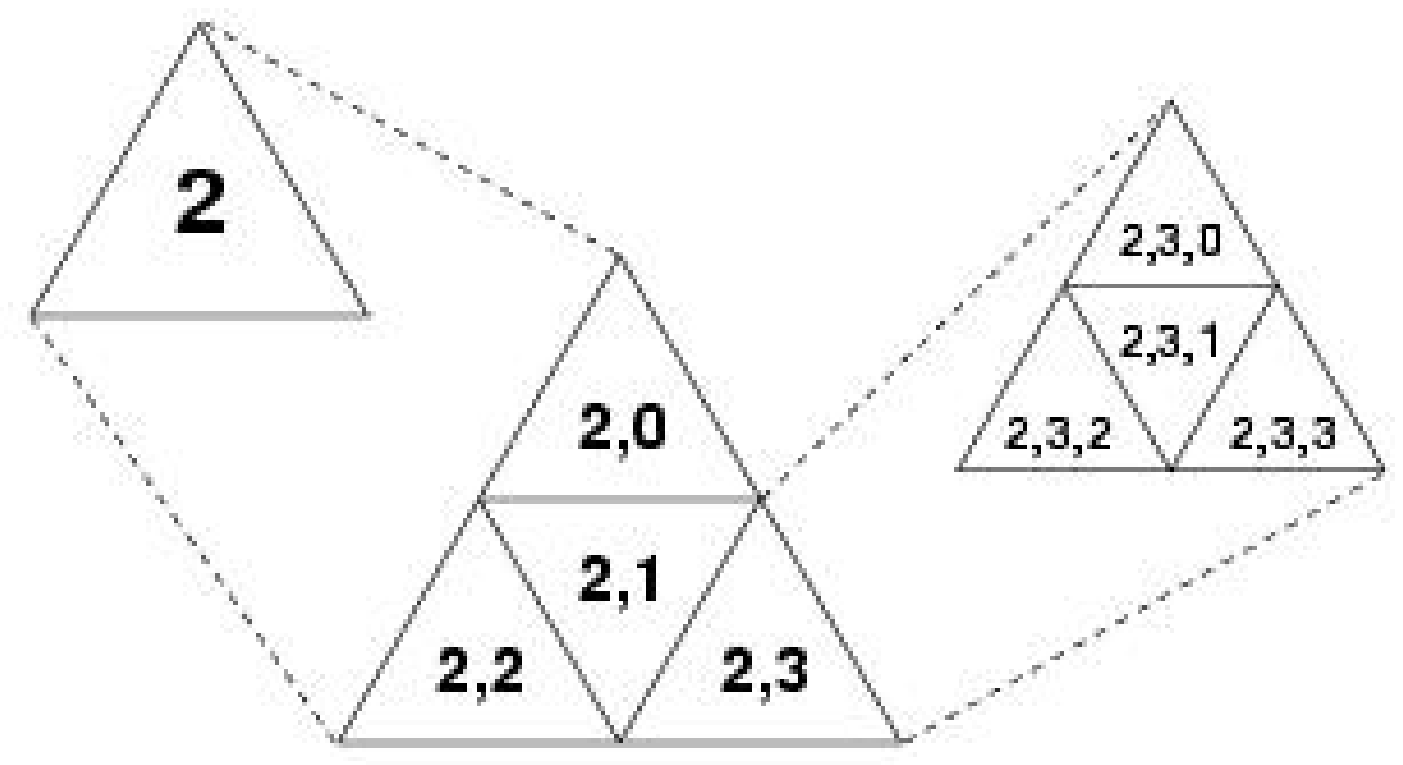}
   \end{tabular}
   \end{center}
   \caption[example] 
   { \label{fig:htm} The idea of the Hierarchical Triangle Mesh
(after Kunszt et al.). A recursive procedure assigns a number to
any point on a sphere. Spatial queries use this index to limit
searches to a relatively small number of triangles with minimum possible area.
}	
   \end{figure} 

\subsection{Logical Database Design} 
\label{sec:logical}

Logical design is concerned with the general data model and data structures
for the database, independently of any hardware issues or the selected
Database Management System (DBMS).
An Object Oriented model is natural wherever there is a need for grouping
parameters and their transformations together (classes). However, performance
problems were reported for a particular Object Oriented Database Management
System (OODBMS) (Ref.~\citenum{Szalay02} and references therein). A relational
system proved more successful in that case. We adopt a relational data model
(see Ref.~\citenum{Connolly01} for a very accessible presentation
of theory). The problem of constructing a database of temporal variability has certain
similarities to a so called data warehouse. It resembles a full record of
all customer receipts in a supermarket rather more than inventory database for the
same store. Object parameters and events are recorded and accumulated in
succession, usually without many further updates. This is the major difference
with respect to a transaction oriented system, where concurrent reads and updates of
previous values are essential. Instead of fluctuating around a fixed size, the
database content tends to grow at a significant rate. Even if we consider only
a fixed amount of data, the structure of chained events remains.
Fig.~\ref{fig:schema} shows the schema we developed for the ROTSE-I database. With minor
modifications this design will handle many other data sets. The backbone is
a main event table containing all available measurements ({\tt Observation}). Each
measurement is identified by a composite primary key that includes the frame
number and the object number. These identifiers are foreign keys which connect
two smaller tables called dimension tables: {\tt Object}, {\tt Frame}.
Dimension tables are basically entry points into the database. Most queries
start from those surrounding tables and are then read from a portion of the main
photometry table. To maximize the speed at the expense of data redundancy
and update anomalies, the dimension tables are often highly denormalized,
resulting in a star schema. We decided against this approach and expanded
the design into a snowflake schema with no unnecessary data redundancy.

We expect cone searches (a ``circle'' in RA and DEC) to be the most
popular use of the database. For that purpose the {\tt Object} table contains
a Hierarchical Triangle Mesh (HTM) index allowing fast extraction of objects
in requested areas of the sky (Section~\ref{sec:spatial}).
Full photometry can be obtained from {\tt Observation}
table for thus selected objects. A separate table contains information specific
to variable objects and the number of entries in {\tt Variable} table is only
a few percent of the number of all objects. Variables are also flagged
in the {\tt Object} table (we model ``is type of'' relationship with a foreign key
in a subclass table). The fact that any given object can be detected in more
than one field is a significant complication. In such cases there may be
multiple object IDs associated with a single physical object. We take this into account
by providing a list of synonymous pairs of object IDs. The advantage of this
approach is that cross identification of multiple references to the same
object can be revised with very little effort as more is known about the data.
In a similar way we precompute the information on nearest neighbors,
providing a valuable diagnostic tool for blending related problems.
The database objects should be cross identified with other surveys.
{\tt CrossID} table in Fig.~\ref{fig:schema} is just a starting point for
thorough cross referencing of different surveys.

On the other side of the main fact table, the {\tt Frame} table provides the
data on individual measurements and allows selection of photometry based
on exposure parameters and observing conditions. Julian Dates of observations
are also stored there and do not have to be duplicated for all objects.
Position dependent effects like heliocentric time correction are not always required
and can be calculated on the fly. The {\tt Frame} is basically in ``part of''
relationship with {\tt Field} because of the observing protocols in synoptic
surveys. A primary key of {\tt Field} is posted in {\tt Frame} to reflect
this fact. Preprocessing of the data and construction of positional templates
on field by field basis implies association of objects with certain fields
(complications occur in the overlap regions). Ideally, such associations
should be removed and all objects should be treated independently of the field
they came from. In practice, full merging of the data is difficult before
all systematic effects are removed from photometry and possibly astrometry as well.
This is another reason to keep the {\tt Field} table in the database:
diagnostic purposes and tracking systematics.

The detections unidentified with any of the template objects are normally
present in every exposure. We store such measurements in {\tt Orphan}
table for further analysis at a later time. Those detections can be spurious, but
in many cases they are real and of great interest, possibly being associated
with moving objects or transients.

\subsection{Physical Considerations} 
\label{sec:physical}

As the basis of our database system we employ PostgreSQL DBMS
({\tt http://www.postgresql.org}), the most advanced
open source system, which is available for free and has a very relaxed license. PostgreSQL
has numerous object oriented features added on top of a relational system.
Formally it handles unlimited amount of data with some minor restrictions like
the 64 TB maximum table size before the need to split. PostgreSQL has
transaction support for real time applications like tracking/updating states of
interesting objects and alert systems. For a fixed size data set of ROTSE-I
we can take maximal approach to indexing, that is define indexes on all primary
and foreign keys and most of the non-floating-point arguments. It remains
to be seen whether frequent rebuilding a large number of indexes in a system with
real-time updates is still an acceptable overhead. We are developing a
web browser based GUI that will be a primary means of connecting to the
database, both internally at LANL and externally over the Internet.
The user interface and most of the database application layer will be based on
ADOdb ({\tt http://php.weblogs.com/ADODB}), a database connectivity library written in PHP.
ADOdb is very portable as it supports numerous DBMSs on several platforms and all
SQL calls are independent of the particular database system,
provided an appropriate driver is loaded.

The main data storage is currently on two 1 GHz Pentium III RedHat Linux boxes,
each with 512 MB of 133 MHz RAM and 600 GB of raw EIDE disk space. Under RAID 5
this gives 1.1 TB of total available space. Both machines are connected
to a 100 baseT ethernet line. Because of the concerns over security and
continuous availability, eventually the entire database will be fully
replicated and only one of the copies will be accessible from outside the local intranet.

\subsection{Spatial Querying} 
\label{sec:spatial}

Until recently RDBMS typically did not include built in structures
for fast querying of spatial data. Positional searches require such
functionality for clustering studies, correlation functions and quick cross
identification of objects. The solution is to construct an external 2D or 3D index,
store it as a column in the object table, and further index it using
one of the standard methods within the database. Following the SDSS Sky Server
experience\cite{Ochsenbein02}, we adopted the Hierarchical Triangular Mesh
(HTM) code\cite{Kunszt00,O'Mullane01} developed at Johns Hopkins University.
The HTM code employs a set of canonical transformations that project the sphere
onto a surface of the octahedron inscribed inside the sphere. The faces of the
octahedron then become the first 8
partitions in the hierarchy. The next level of the nested structure is
recursively defined by 3 vertices and 3 bisects of each side of the
triangle at any given level. Fig.~\ref{fig:htm} (after Kunszt\cite{Kunszt00})
illustrates the idea. The HTM software package also provides functions that return
a set of triangles covering circular and polygonal areas. With proper indexing
objects inside those triangles can be quickly searched through.

Although designed for slightly different applications, R-trees can also be
adapted for spatial range queries of point objects, and have been reported to
provide acceptable performance\cite{Baruffolo99}. R-tree based indexing
is available in PostgreSQL and could be a viable alternative to HTM.

   \begin{figure}[H]
   \begin{center}
   \begin{tabular}{c}
   \includegraphics[height=9cm]{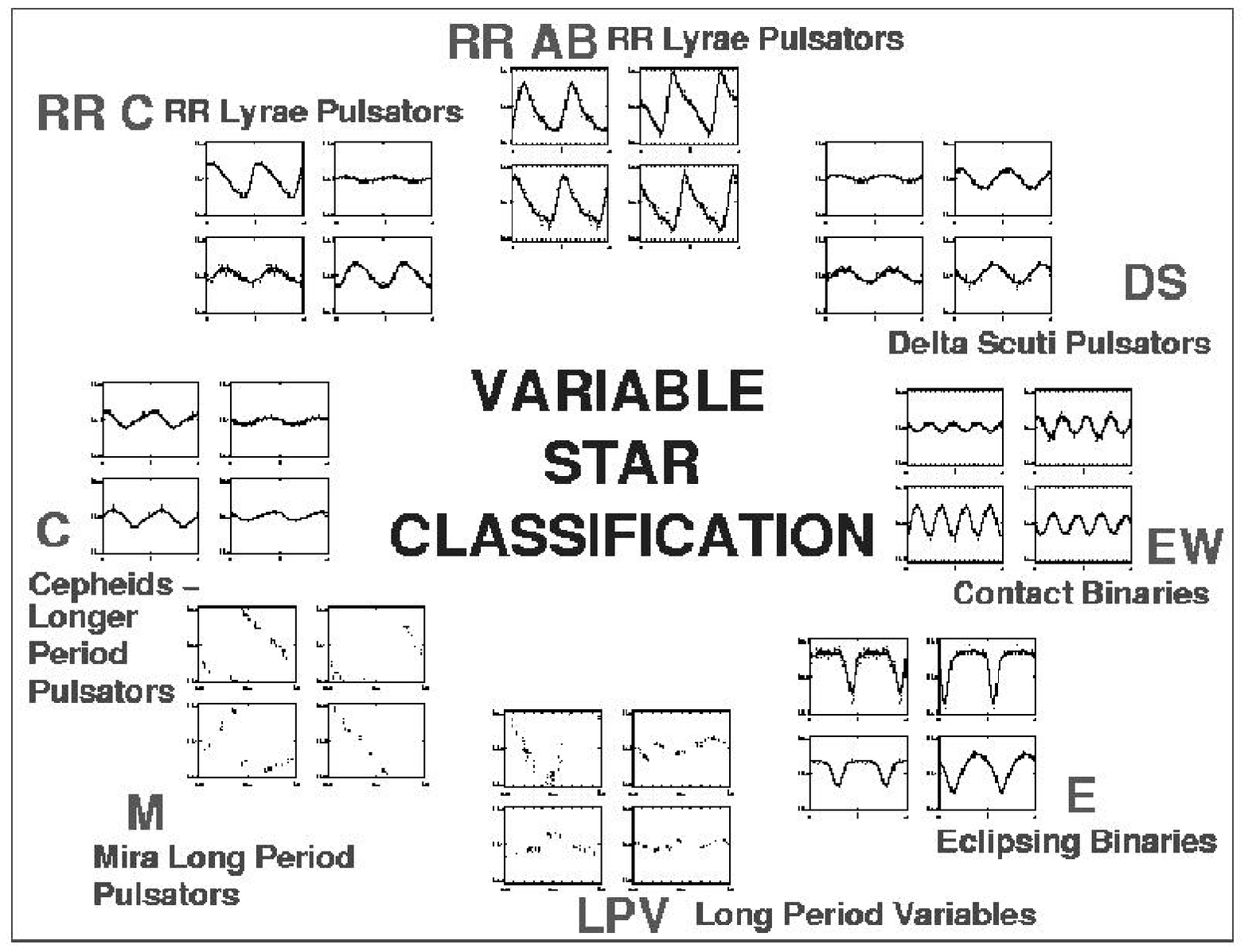}
   \end{tabular}
   \end{center}
   \caption[example] 
   { \label{fig:classification} An illustration of variable star
classification problem from Section~\ref{sec:mining}. This example of
classification is based entirely on light curve parameters
(period, amplitude, and shape).
}
   \end{figure} 

   \begin{figure}[H]
   \begin{center}
   \begin{tabular}{c}
   \includegraphics[height=9cm]{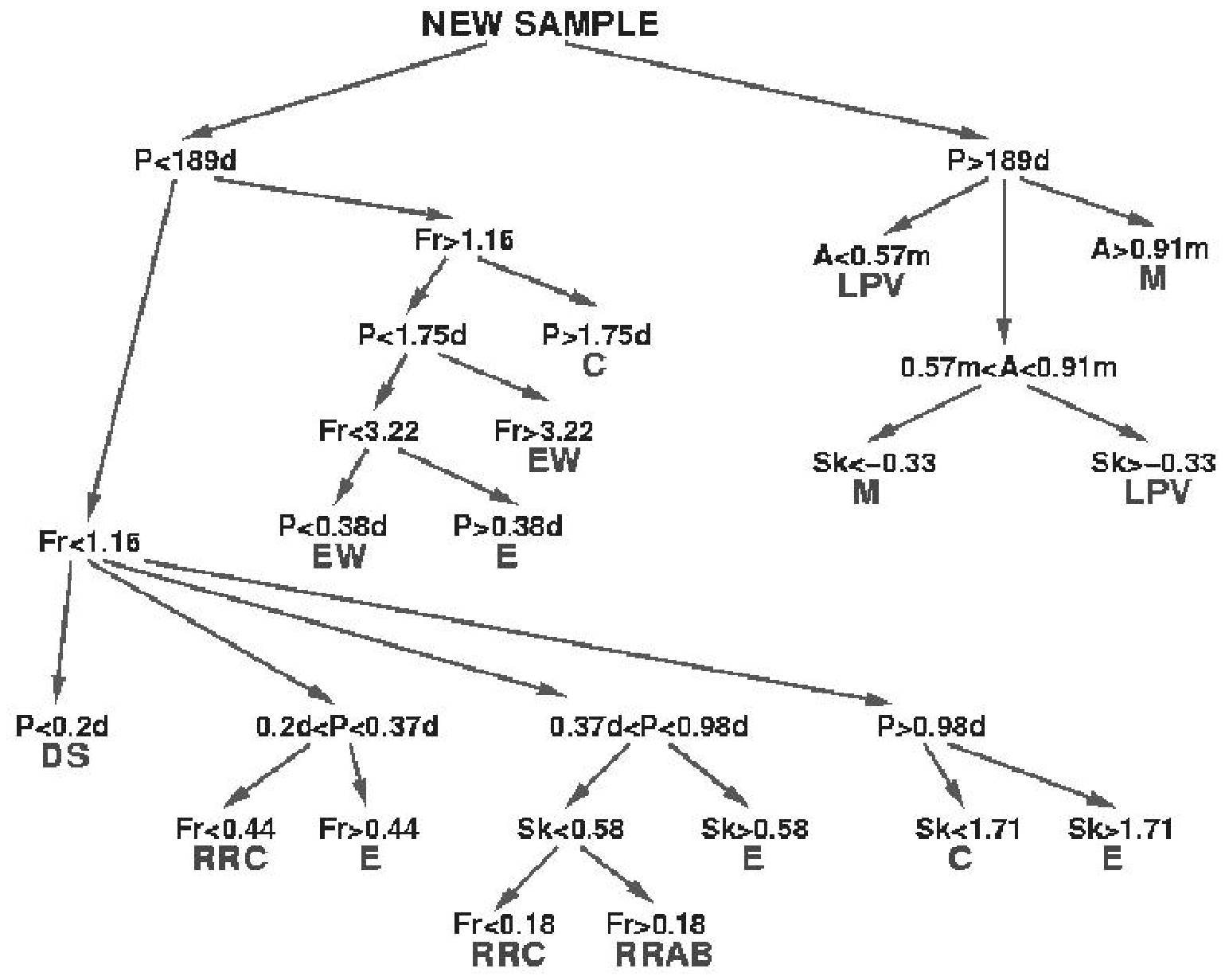}
   \end{tabular}
   \end{center}
   \caption[example] 
   { \label{fig:tree} A binary tree obtained with the J4.8 algorithm
for classification in Section~\ref{sec:mining}.
The main characteristics of the human classification are reproduced:
e.g. period ranges at the bottom. The complexity of the tree was
reduced using sub-tree rising by requiring a minimum 10 objects
at any given leaf node and 90\% confidence for pruning.
}
   \end{figure} 

\section{Data Mining}

The ultimate goal behind our efforts is to provide easy access to various
variability data in order to enable extraction of new astronomical information
for research on source variability.
The web interface under construction will provide a set of high level tools
for data analysis and visualization. Our strategy is to start from very basic
functionality and gradually build a powerful data mining system. Ultimately,
after extracting the required data, the user will have an option to run period
searches, phase the data with an arbitrary period, perform other time series
analysis, i.e., correlation functions, Fourier transforms and smoothing.
The interface will allow users to plot various database
stored and derived parameters, display sky maps with overlayed
selection of objects, cross correlate various data sets and finally classify
objects using both basic information as well as the information obtained in the
course of the session. In the area of classification and clustering, Machine
Learning has made remarkable progress in recent years. Given the growing complexity of
astronomical data (high dimensionality with dimensions describing very diverse
and sophisticated characteristics), mining large databases will require new
efficient tools. The results from Machine Learning are typically more
objective, since this approach often eliminates human bias and error.

   \begin{figure}
   \begin{center}
   \begin{tabular}{c}
   \includegraphics[height=9cm]{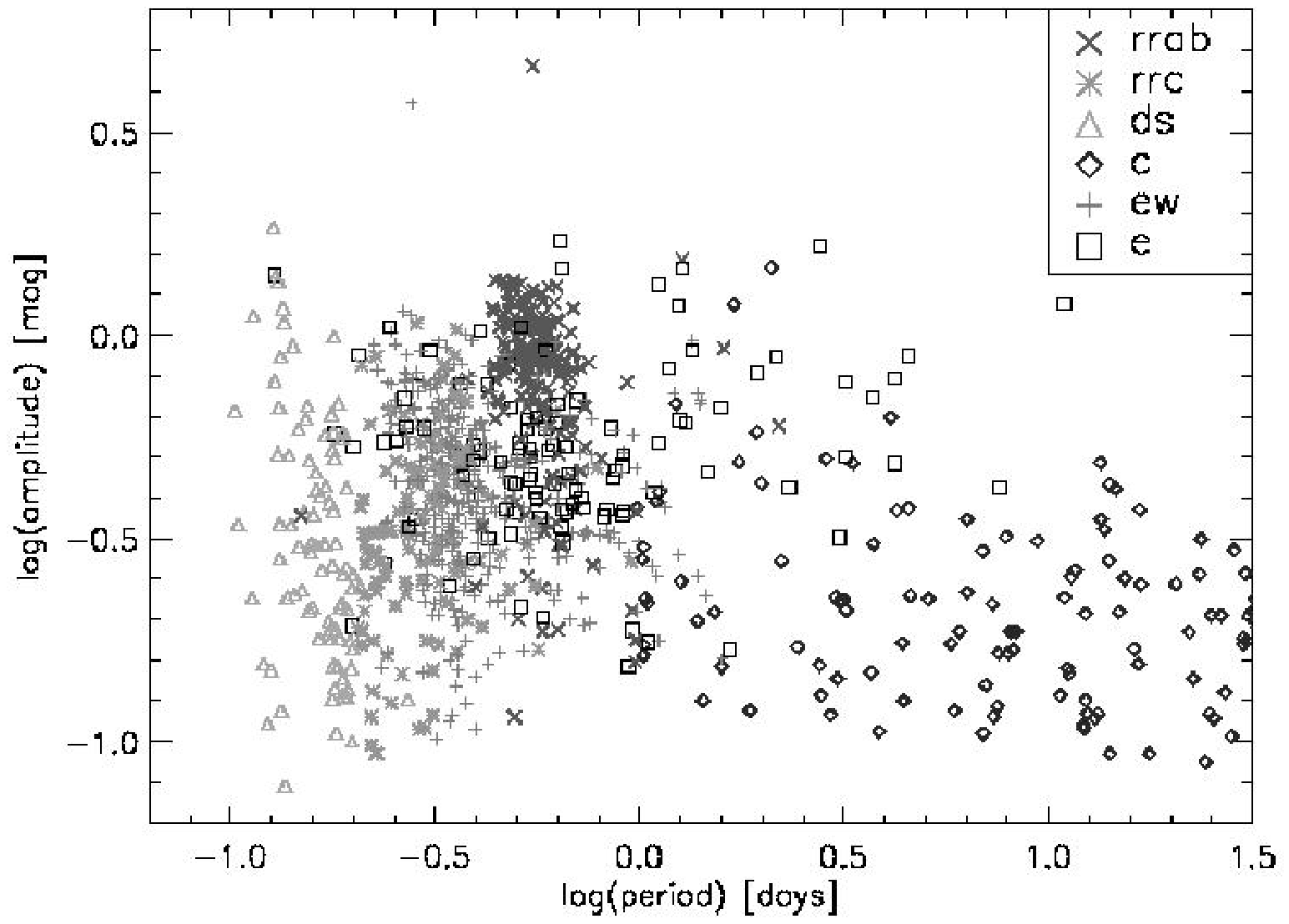}\\
   \includegraphics[height=9cm]{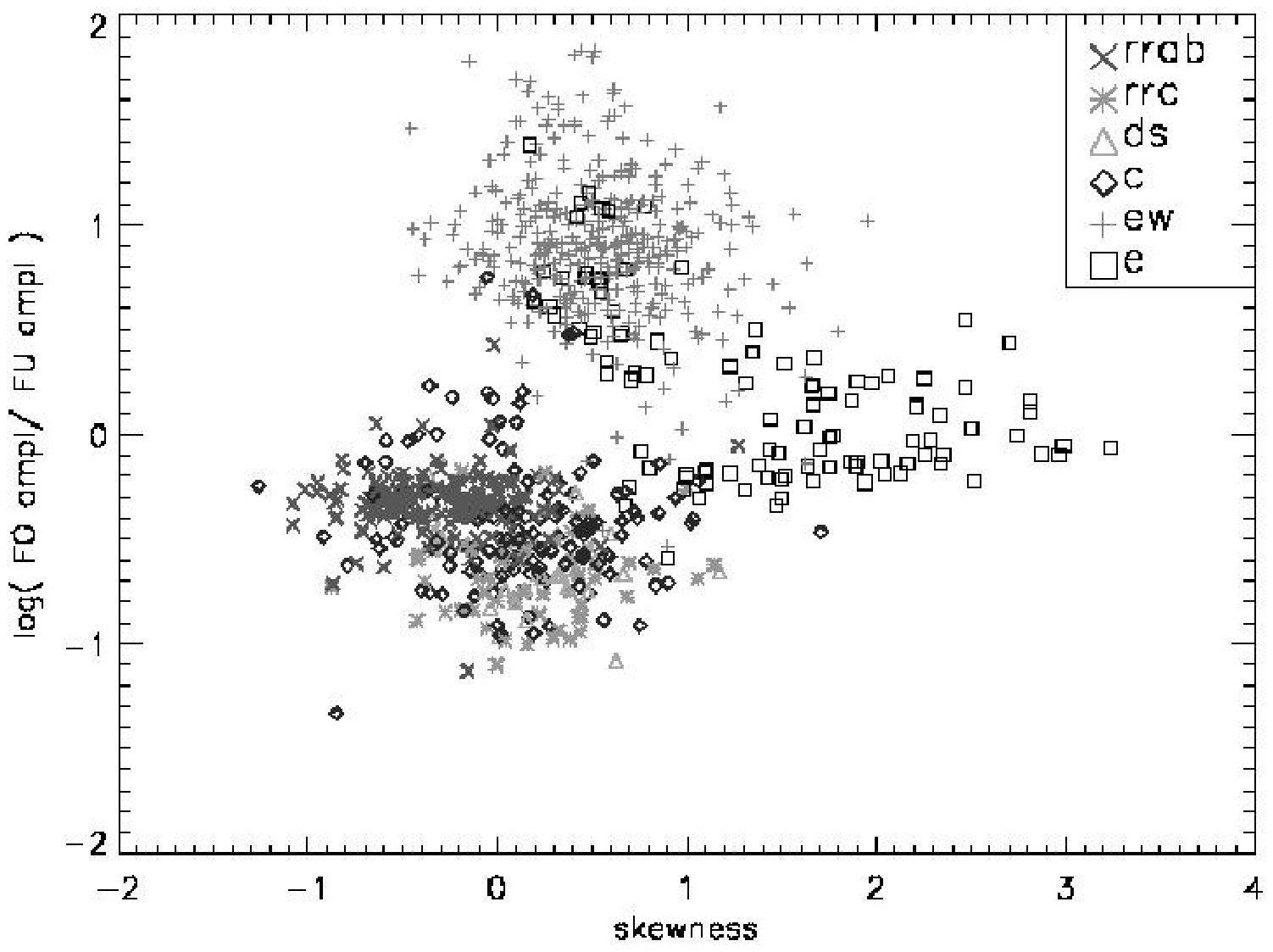}
   \end{tabular}
   \end{center}
   \caption[example] 
   { \label{fig:features} Feature space for our example of variable star
classification. For a sample of 1700 periodic variables two projections are
shown: Period-Amplitude (upper) and Skewness-Fourier ratio (lower).
}
   \end{figure} 

\begin{table}[h]
\caption{Results for variable star classification in Section~\ref{sec:mining}.}
\label{tab:results}
\begin{center}       
\begin{tabular}{|l|r|c|c|} 
\hline
\multicolumn{2}{|c|}{~~} & \multicolumn{2}{|c|}{Accuracy} \\ 
\cline{3-4}
\multicolumn{2}{|c|}{Method} & cross-validation & training \\
\multicolumn{2}{|c|}{~~} & data & data \\
\hline
\rule[-1ex]{0pt}{3.5ex} Supervised &  SVM & 90\% & 95\% \\
\cline{2-4}
\rule[-1ex]{0pt}{3.5ex}  ~~        & J4.8 & 90\% & 92\% \\
\cline{2-4}
\rule[-1ex]{0pt}{3.5ex}  ~~        & 5-NN & 81\% & 86\% \\
\hline
\rule[-1ex]{0pt}{3.5ex}  Unsupervised &  Autoclass & 80\% & 80\% \\
\cline{2-4}
\rule[-1ex]{0pt}{3.5ex}  ~~           &  k-Means   &  --- & 70\% \\
\hline
\end{tabular}
\end{center}
\end{table}

\subsection{Classification of Variable Stars} 
\label{sec:mining}

Currently we are evaluating machine learning algorithms for potential use in
classification of astronomical objects. Two basic types of methods can be
distinguished. Supervised techniques require a training sample with known
class membership. Unsupervised learning, on the other hand, works on data
with unknown classifications, including an unknown number of classes for some
algorithms. Supervised methods naturally outperform unsupervised ones, however
the use of training sample with known classifications imposes prior knowledge
onto the final result and may not always be possible. Unsupervised algorithms
are capable of finding entirely new classes of objects.

Using several supervised and unsupervised techniques we classified a set of
$\sim1700$ periodic variable stars from a pilot search of the ROTSE
data\cite{Akerlof00b}. The data has been previously classified with the
human made algorithm based on prior knowledge and the appearance of the scatter
plots. In the study of the pilot ROTSE sample the labels have been verified by
visual inspection of light curves. In Fig.~\ref{fig:classification} we show
prototype objects for 8 classes out of 9 considered in our study (with the
exception of the catch-all class ``other'').
The location of objects in parameter space is shown in
Fig.~\ref{fig:features}. Classification is based on a light curve in a single
photometric band only (period, amplitude, the ratio of first
overtone to fundamental frequencies and the skewness of the magnitude
distribution).

Table~\ref{tab:results} summarizes our results for Support Vector Machines
(SVM)\cite{Vapnik98}, decision tree builder (J4.8), five nearest neighbors
classifier (5-NN), k-means clusterer\cite{Witten99} and Bayesian system
Autoclass\cite{Cheeseman96}. More details on SVM tests on variable star
data can be found in Ref.~\citenum{Wozniak02c}.
Performance on training data is usually a poor predictor of the performance on
future data. Much better results can be obtained from cross validation
analysis. We performed a 5-fold cross validation, where  randomly selected
4/5 of the sample is used for training and then the accuracy is evaluated on
the remaining 1/5. With the state of the art SVM method, we achieved 90\%
accuracy for the full problem (9 classes). It was estimated that two human
analysts would agree at a similar level. Actually for a fraction of cases,
even the same person making the same classification twice, arrives at
a different conclusion each time.
Performance can be as good as 95 or even 98 \% for a restricted problem when
larger, more general classes are considered or when objects of one class are
detected against the background of ``everything else''. We have noted a very
good potential for the use of decision trees. This type of algorithm is
particularly attractive because it gives full insight into how the features
were used to compute the classification, and the machine can then convey the
knowledge back to a human. The tree in Fig.~\ref{fig:tree} correctly reproduces
many of the features found in human made algorithm, despite the fact that
we worked with a feature space somewhat different from the one used
in the benchmark visual study. The tree was trimmed of the least populated
branches and leaves to control over-fitting. We kept the final accuracy
of the tree at 90\%.

\section{Summary and Discussion}

There are considerable scientific rewards coming out of massive data sets in
astronomy. SkyDOT, a general sky variability database being designed and
constructed at LANL, should have a major beneficial influence on the field.
Several surveys either conducted locally or with indirect LANL involvement will
provide the data for the database. The future synoptic data sets will be
integrated into the database to maximize the scientific potential. The current
activities focus on data conversion and web database application software.
The first results in application of Machine Learning to this type of data
are promising. Data Mining technologies will eventually be integrated with the
database and may aid new surprising discoveries. The National Virtual
Observatory bears the promise that astronomical community will be prepared for
challenges in data overloaded astronomy. Recently the NVO, in collaboration
with similar European initiatives, announced VOTable\cite{Williams02},
a new XML based standard for exchanging astronomical data\cite{Ochsenbein02}.
We are closely following activities in this area to assure NVO compliant output
from our database.

\acknowledgments     
 
This work is supported by the Laboratory Directed Research and Development
funds at LANL under DOE contract W-7405-ENG-36.

\clearpage


\end{document}